\documentclass[usegraphicx]{mn2e}

\usepackage{times}

\title[SMC red variables]{Red variables in the
OGLE-II database. II. Comparison of the Large and Small Magellanic
Clouds}
\author[L. L. Kiss and T. R. Bedding]{L. L. Kiss\thanks{E-mail:
laszlo@physics.usyd.edu.au}\thanks{On leave from University of Szeged,
Hungary} and T. R. Bedding\\
\\
School of Physics, University of Sydney 2006, Australia}

\begin{document}

\date{Accepted ... Received ..; in original form ..}


\maketitle

\begin{abstract}

We present period-luminosity relations for more than 3,200 red variable stars in
the Small Magellanic Cloud observed in the second phase of the Optical 
Gravitational Lensing Experiment (OGLE-II). Periods of multiply-periodic light 
curve solutions combined with the single-epoch 2MASS $JHK_{\rm S}$ magnitudes, 
reveal very similar distributions to those for the Large Magellanic Cloud 
in Paper I. The main features include four pulsating Asymptotic Giant Branch
(AGB) ridges, three distinct 
short-period sequences below the tip of the Red Giant Branch and two long-period
sequences of ambiguous origin. We derive a relative distance modulus of the 
Clouds from the period-luminosity distributions for all stars of 
$\Delta \mu_0=0.44$ mag, which is in good
agreement with recent independent results. The tip of the Red
Giant Branch shows a colour and metallicity dependence that is in 
excellent agreement with the empirical results for globular clusters.
We conclude that most variable stars below the TRGB are indeed RGB stars.
\end{abstract}

\begin{keywords}
stars: late-type -- stars: variables -- stars: oscillations -- stars: AGB and
post-AGB 
\end{keywords}

\section{Introduction}

The period-luminosity (P-L) relations of long period variables on the AGB
are well documented for the Large Magellanic Cloud. Observations have revealed
not only that Mira variables follow a tight near-infrared P-L relation 
(Glass \& Lloyd
Evans 1981, 2003; Feast et al. 1989), but also that semiregular variables
pulsating in overtone modes follow distinct and well-defined P-L relations (Wood
et al. 1999, Wood 2000). In recent years, there has been strong interest in
pulsating red giants, which is reflected in the increasing number of independent
analyses  of large photometric databases (Cioni et al. 2001, 2003, Noda et al.
2002,  Lebzelter et al. 2002, Glass \& Schultheis 2003, Kiss \& Bedding 2003, 
Ita et al. 2003, Groenewegen 2004). 

An interesting by-product was the recognition  of a distinct and very numerous
group of red variables below the tip of the  Red Giant Branch (TRGB). First, Ita
et al. (2002) noted that the luminosity function (LF) of red giant variable stars
in the LMC shows a sharp feature at the expected brightness of the TRGB. Earlier,
Alves et al. (1998) and Wood (2000) suggested that thermally pulsing AGB stars
can mimick a two-peaked luminosity function, and thus the LF shape alone is not
sufficient to establish the  existence of TRGB variables. However, we have shown
in Paper I (Kiss \& Bedding 2003) that the LMC period-luminosity relations for
stars above and below the TRGB show a relative period shift that is consistent
with evolutionary temperature difference between RGB and AGB stars.  This result
has since been confirmed by Ita et al. (2003). The possibility of observable
pulsations in RGB stars therefore  seems to be well  established, providing a new
area for asteroseismological considerations. In this Letter we provide further
evidence for extragalactic RGB variables through a comparative analysis of red
variables in the Small Magellanic Cloud (SMC). 

Red giant pulsators in the SMC have a far less  extensive literature than those
of the LMC. The largest pre-microlensing datasets for the SMC were published by
Wood et al. (1981), Lloyd Evans et al. (1988) and Sebo \&  Wood (1994), all
consisting of a few dozen variables at most. Microlensing surveys (MACHO, EROS,
OGLE) have changed the situation by discovering thousands of red variables.
Based on the extensive new datasets, it has became possible to compare 
the Magellanic Clouds and the Galactic Bulge in search for metallicity effects
on red giant pulsations and P-L relations. 
Cioni et al. (2003) examined MACHO light curves of 458 LPVs in the  SMC that 
were detected by the Infrared Space Observatory. They found very similar  P-L
sequences to those in the LMC and confirmed the relative overabundance of 
carbon-rich Mira stars (as recognized first by Lloyd Evans
et al. 1988). The most extensive analysis to date is that of Ita et al. (2003),
who determined the dominant  period for 2,927 variables in the SMC using OGLE-II
data and constructed  P-L relations using single-epoch SIRIUS $JHK$ survey data.
Beside finding the same distributions as those in Paper I, Ita et al. (2003)
derived zero-point differences of the P-L relations of about 0.1 mag. Some
differences were also noted recently by  Cioni et al. (2003) for the Magellanic
Clouds and  by Glass \& Schultheis (2003) for the Galactic Bulge and the
Magellanic Clouds. The latter investigators found both slightly different slopes
and  different absolute magnitude ranges for the overtone P-L sequences. 
These interesting results depend on the assumed distances to each of 
the different P-L datasets. Here we concentrate on the multi-wavelength 
luminosity functions to sort out the distance issue, and examine the 
implications concerning the RGB pulsations.

\section{Data analysis}

The basis of our analysis is the catalogue of OGLE-II (Udalski et al. 1997) 
data spanning four years 
from 1997 to 2000 (Zebrun et al. 2001). Of the total 7 square degrees  of the
OGLE Magellanic Cloud fields, about 2.5 square degrees were observed  in the
SMC. The observations were analysed by the OGLE team using a modification of
the Difference Image Analysis (Alard \& Lupton, 1998, Wozniak 2000), which is
the most advanced method available for detecting small variations in crowded
fields (see, e.g., Bonanos \& Stanek 2003). 

Typical light curves in the SMC consist of 280 points distributed  over
$\sim$1,100 days. Compared to the LMC, where individual datasets covered
slightly more than 1,200 days in 400 points, this means a reduced accuracy in
period determination. We consider our periods reliable up to  about 400--500
days, for which at least two cycles were observed. 

The data  were analysed in the same manner as in Paper I.  To summarise, we
performed the following steps:

\begin{enumerate}

\item We cross-correlated the 15,038 OGLE-II variable stars in the SMC
with the 2MASS All-Sky Point Source Catalog\footnote{
\tt http://irsa.ipac.caltech.edu}; within a search radius of $1^{\prime\prime}$
we found 10,361 stars with $JHK_{\rm S}$ magnitudes. The tight constraint on
coordinates ensured that we identified the infrared counterparts 
unambiguously.

\item After excluding duplicates ($<$10), we selected red giants according to
the $J-K_{\rm S}$ colour. The applied threshold was $J-K_{\rm S}>0.9$ mag. 
We have also excluded all stars with $K_{\rm S}>14$ mag 
because of the strongly increasing errors of 2MASS magnitudes below 
that limit. The final sample consisted of 3,898 stars.

\item We performed a period search on OGLE-II light curves by an iterative 
Fourier analysis (see Paper I for details), until a four-component harmonic fit
was calculated. We kept only those frequencies larger than $8 \times 10^{-4}$
d$^{-1}$ ($\sim 1/T_{\rm obs}$) and with semi-amplitudes larger than 5
mmag.

\item The resulting database contains 3,260 stars with 10,009 periods (and
amplitudes and phases), the mean $I$-band magnitude and 2MASS $JHK_{\rm s}$
single-epoch magnitudes.

\end{enumerate}

Our sample is slightly larger than that of Ita et al. (2003) because of the
different selection criteria. A more relevant difference is that we calculated a
set of periods for every star, while Ita et al. (2003) determined only one period
per star. We found multiple periodicity for the majority of stars (82\% of
the full sample), which can, therefore, be identified as common behaviour in red
giants. There must
be, of course, a  fraction of spurious or physically unrelated periods, which
need a more sophisticated method for identification and exclusion from future
analyses. 

\section{Discussion}

\begin{figure}
\includegraphics[width=80mm]{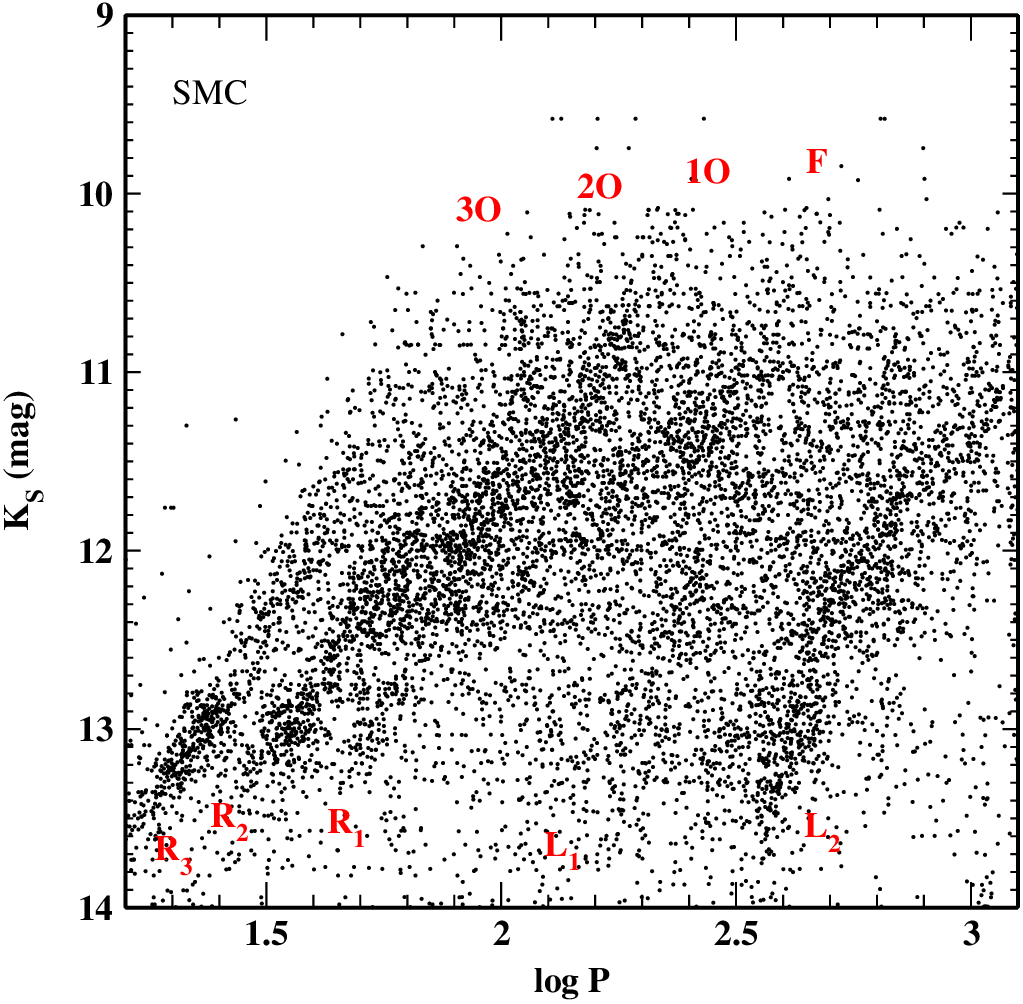}
\caption{P-L relations for red giants in the SMC 
(10,009 periods for 3,260 stars).}
\label{plr}
\end{figure}

We present the resulting P-L diagram in Fig\ \ref{plr}.
At first glance, the shape and extension of distributions are very similar to
those we found  for the LMC: {\it (i)} there is a noticeable drop of point
density at $K_{\rm s}\approx12.70$ mag, located close to the tip of the RGB
(Cioni et al. 2000); {\it (ii)} above the tip, there are four pulsating AGB
sequences, which in the LMC were identified by Wood (2000) as stars pulsating
in the fundamental (F), first (1O), second (2O) and  third (3O) overtone modes
(he labelled them as sequences C, B and A which, however, does not account for
the fact that there are four ridges clearly distinguishable in the OGLE-II
data); {\it (iii)} below the tip,  there are three distinct sequences in the
short-period range ($P<60$ days), of which  R$_2$ and R$_3$ are continuations
of 2O and 3O; there might be a slight horizontal shift, similar to that found
in the LMC  (Paper I), but the ridges are less well defined, due to the 
smaller number of stars, 
so the possible period shift between 3O -- R$_3$ and between 2O -- R$_2$ is hardly
measurable. The sequence R$_1$, just as for the LMC, lies somewhere between the
extrapolated 1O and F ridges; {\it (iv)} for longer periods, there are two
sequences, L$_1$ and L$_2$, in the same relative position to the other ridges
as in the LMC. Their interpretation is ambiguous; Wood (2000) suggested L$_1$
(=E in his paper) contains eclipsing binary variables, while L$_2$ (=D) may consist
of pulsating stars with unknown excitation mechanism or non-spherical stars
with rotationally induced  variatons (Olivier \&  Wood 2003, Wood 2004). 

Although multiperiodicity is a common feature, 
there is a remarkable difference above and below the tip of the RGB: 
for $K_{\rm S}<12.7$ mag, 95\%, while for $K_{\rm S}>12.7$, only 62\% of 
the light curves resulted in multiple periods. This, however, only reflects the
fact that amplitudes below the TRGB strongly decreases towards fainter magnitudes 
(see Sect. 3.3), thus the fixed 5 mmag amplitude threshold may be too high 
for most of the periods (if they exist). In addition, we stress there is not 
any noticeable change between the resulting P-L distributions when plotting every 
star just once using the strongest period or using all periods with amplitudes
exceeding a certain threshold (or even plotting every star just once with the 
second strongest period).

\subsection{The relative distance modulus of the Clouds}

In the case of such complex structures as the P-L distributions of red variables
in the Magellanic Clouds, it is difficult to measure the vertical distance
between  the P-L ridges directly. It has been a more common approach to assume a
relative distance modulus  and overplot the relations with the purpose of
comparison, either in a simplified functional form (see Fig.\ 11 in Glass \&
Schultheis 2003) or showing the full observational data (Fig.\ 10 in Ita et al.
2003). Here we follow a different approach to show that, assuming a negligible
mean zero-point difference for the set of relations, one can get consistent
relative distance moduli and extinctions of the Clouds in all four photometric
bands ($IJHK_{\rm S}$), which will be used later to characterize the dips of the
luminosity functions.

\begin{table}
\begin{center}
\caption{Relative distance moduli and extinctions. $\Delta \mu$ is the mean
vertical distance between the P-L images; $\Delta A_X$ refers to the 
differential extinction in 
the $X$ band, while $\Delta \mu_0$ is the dereddened relative distance modulus.
$\Delta E(B-V)=0.085$ (Westerlund 1997) and $\Delta A_V=0.26$ mag were assumed.
The uncertainty in $\Delta \mu$ is about $\pm$0.05 mag.}
\label{deltamu}
\begin{tabular}{|lccc|}
\hline
Band & $\Delta\mu$  & $\Delta A_X$ & $\Delta\mu_0$\\
\hline
$I$  & 0.25 & 0.16 & 0.41\\
$J$  & 0.35 & 0.07 & 0.42\\
$H$  & 0.40 & 0.05 & 0.45 \\
$K_{\rm S}$ & 0.43 & 0.03 & 0.46\\
\hline
mean: & & &0.44\\
\hline
\end{tabular}
\end{center}
\end{table}

Instead of a visual examination of overplotted data, we calculated the
cross-correlation functions of smoothed images of the P-L ridges (being the
sharpest in $K_{\rm S}$ and the least-defined in $I$). For this, we converted the
scatter diagrams (Fig.\ \ref{plr} and Fig.\ 1 in Paper I) to $500\times500$ pixel
images as follows. The x-axis covered $\log P$ between 1.2 and 3.1  
($3.8\times10^{-3}$ dex/pixel), while the y-axis covered 5 magnitudes, where the
exact range depended on the studied band (e.g., from 9  mag to 14 mag in $K_{\rm
S}$). This way one pixel in vertical direction corresponds to 0.01 mag. Every
pixel ``intensity'' value was set to the total number of stars in a $21\times21$
pixel mask, centered on the actual pixel. This masking worked as a boxcar
smoothing procedure, which produced clear images of the P-L ridges. The
cross-correlation functions of the $IJHK_{\rm S}$ P-L images were calculated as
functions of $\Delta\mu$, and their maxima showed the vertical shifts that give
the best overlap of the images. In order to minimize the effects of spurious
periods, we kept only stars with $\log P<$\ 2.6 (i.e. $P<$\ 1 year); in other
words, we used only $360\times500$ pixel sub-images.

The resulting shifts are presented in the second column of Table\ \ref{deltamu}.
There is a strong wavelength dependence, which can be explained by the
differential reddening. The average $E(B-V)$ for the LMC and the SMC are 0.15 mag
and 0.065 mag, respectively (Westerlund 1997), which can be translated to $\Delta
A_X$ differential extinction by adopting $R_V=3.1$ and the relative extinction 
coefficients ($A_X/A_V$) in Schlegel et al. (1998). 

As can be seen in Table\ \ref{deltamu}, the dereddened relative distance moduli
agree well within the error bars ($\sim\pm$0.05 mag in every band). Similar
agreement exists for recently published results; for instance, Cioni et al. (2000)
determined $\Delta \mu=0.44$ mag. On the other hand, Ita et al. (2003) adopted 
$\Delta \mu=0.49$ mag, while Glass \& Schultheis (2003) used 0.50 mag. In the rest
of the paper we adopt 0.44 mag with an estimated $\pm$0.03 mag random error.

\subsection{The luminosity functions and the tip of the RGB}

\begin{figure}
\includegraphics[width=80mm]{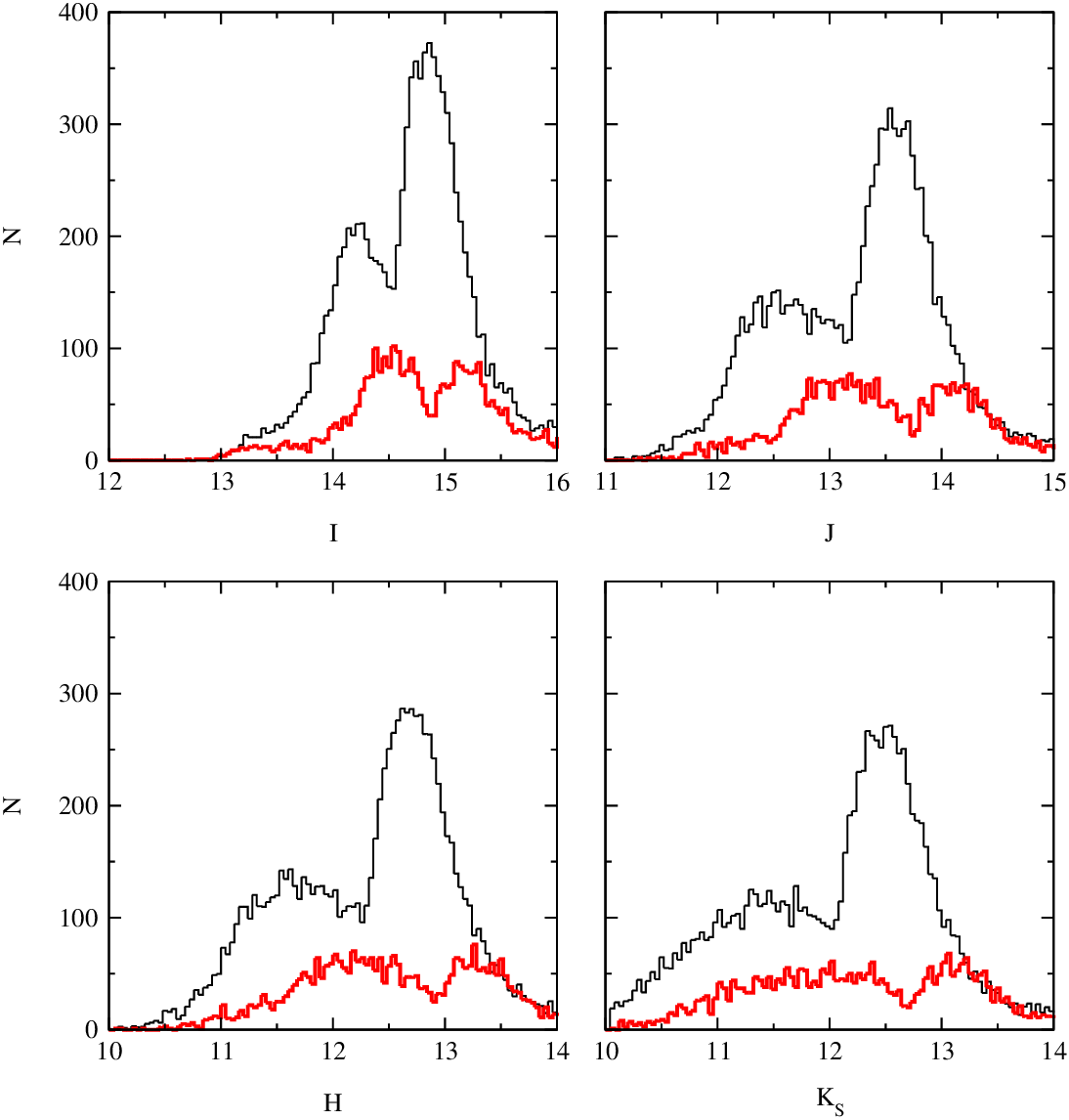}
\caption{Stellar magnitude distributions for OGLE-II variables
in the SMC (thick red line) and in the LMC (thin black line), normalized by the
survey area.}
\label{lfs}
\end{figure}

\begin{figure}
\includegraphics[width=80mm]{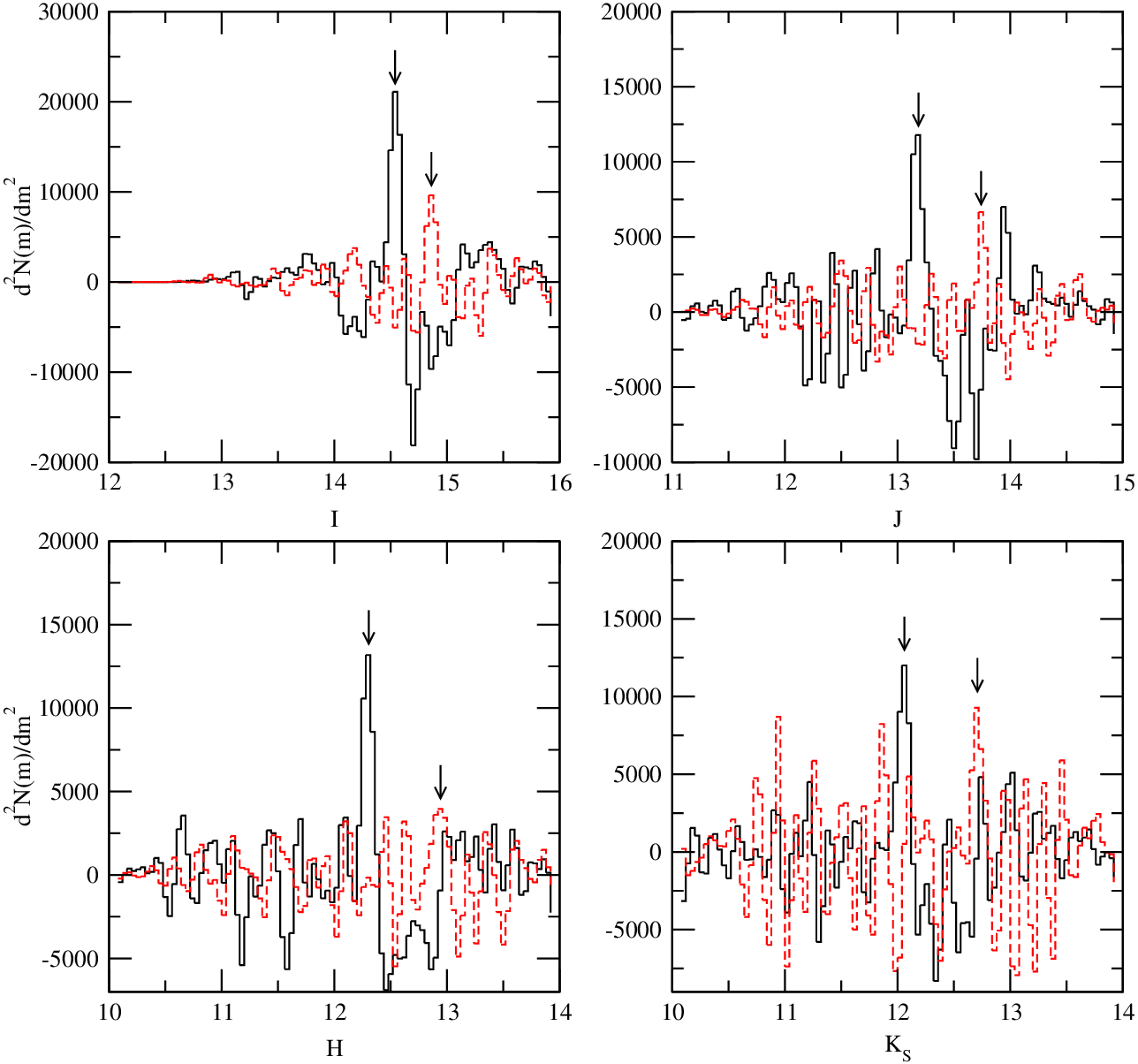}
\caption{The second derivatives of the smoothed luminosity functions. The arrows
point to the maxima for the LMC (solid lines) and the SMC (dashed lines).}
\label{trgbs}
\end{figure}

In Fig.\ \ref{lfs} we plot luminosity functions in all four bands of the
OGLE-II+2MASS database: the thin solid line shows the LMC data after correction
for the different survey  areas (4.5 square degrees for the LMC vs.
2.5 square degrees for the SMC); the thick solid line denotes the SMC LF.

Both Clouds have double-peaked LFs with similar shapes and wavelength 
dependences. The exact overlap of the faint ends of the SMC and
area-normalized LMC distributions in $JHK_{\rm S}$  shows that there is a severe
sensitivity cut-off for $J>14.2$ mag, $H>13.5$ and $K_{\rm S}>13.3$ mag. On the
other hand, the difference for $I$ suggests the drop of the OGLE-II detection
efficiency occurs at fainter magnitudes, so that the faint limit in our 
sample is set by 2MASS and not by OGLE.

An interesting point is the changing shape of the AGB peak. It is the
narrowest in $I$ (where the full-width-at-half-maximum is 0.7 mag)  and going
toward the  redder bands, the stars spread over wider and wider luminosity
ranges  (the FWHM increases to 2 mag). This can be understood qualitatively 
in terms of the larger bolometric correction for the brightest and
reddest stars (Alvarez et al. 2000), which results in the observed narrower
distribution in $I$. We point out the overall similarity of the two AGB LFs, with
only minor differences.

\begin{figure*}
\includegraphics[width=160mm]{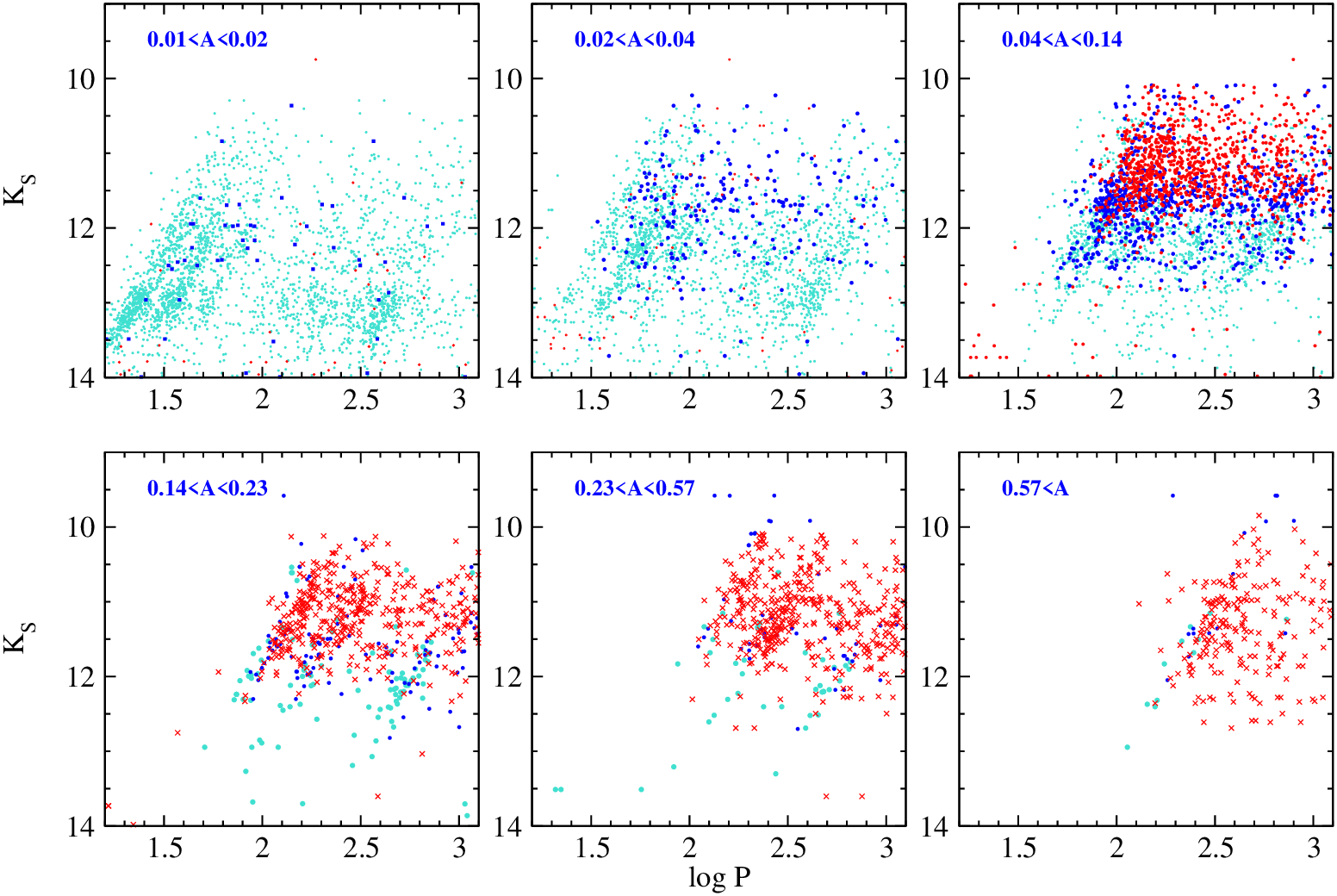}
\caption{P-L relations in the SMC as function of the full amplitude of modes.
Three $J-K_S$ colour ranges were selected to plot in different colours.
Turquoise (light gray): $0.9<J-K_S\leq1.2$; blue (black): $1.2<J-K_S\leq1.4$; 
red (dark gray): $J-K_S>1.4$. This figure is available in colour in the on-line
version of the journal on {\it Synergy}. Symbols in
the lower panels are drawn larger for improved clarity.}
\label{plra}
\end{figure*}

The fainter peaks, on the other hand, show quite different behaviour. The
FWHM stays essentially constant with wavelength, with slightly narrower
distributions for the SMC. Since the RGB luminosity function is increasing
monotonically toward fainter magnitudes (Nikolaev \& Weinberg 2000), the sharp
cut-off in our data is a consequence of the falling detection efficiency in
the combined OGLE-II+2MASS database. The dip between the brighter
and fainter peaks is very close to the TRGB magnitudes determined by Cioni et
al. (2000) from an analysis  of about 150,000 objects in the DENIS catalog and
that close proximity was interpreted both by Ita et al. (2002) and us in
Paper I as the evidence for pulsating RGB stars.

\begin{table}
\begin{center}
\caption{Summary of TRGB magnitude values (the uncertainty is about 0.04 mag). 
C2000 refers to TRGB values in Cioni et al. (2000).}
\label{mtrgb}
\begin{tabular}{|lcccccc|}
\hline
Band & LMC & SMC & LMC & SMC & LMC & SMC\\
     & LF dip & LF dip & $f^{\prime\prime}_{\rm obs}$ & $f^{\prime\prime}_{\rm
     obs}$ & (C2000) & (C2000)\\
\hline
$I$  & 14.56 & 14.92 & 14.54 & 14.85 & 14.54 & 14.95\\
$J$  & 13.20 & 13.79 & 13.18 & 13.74 & 13.17 & 13.73\\
$H$  & 12.27 &  12.90 & 12.30 & 12.94 & -- & -- \\
$K_{\rm S}$ & 12.03 & 12.72 & 12.06 & 12.70 & 11.98 & 12.62\\
\hline
\end{tabular}
\end{center}
\end{table}

Following the referee's recommendations, we examined this agreement more closely.
Due to the nature of the tip of the RGB, it manifests as an edge in
the luminosity function ($f_{\rm obs}$). Cioni et al. (2000) have shown 
very thoroughly in their
Appendix that making an unbiased estimate of the discontinuity 
can be very difficult. Previous authors generally determined the position of the
peak in $f^\prime_{\rm obs}$, while Cioni et al. (2000) argued for using
$f^{\prime\prime}_{\rm obs}$. Adopting their considerations, we determined  the
second derivatives of smoothed luminosity functions in all bands (Fig.\
\ref{trgbs}). The OGLE-II LFs were smoothed with a Gaussian weight-function 
(FWHM=0.04 mag) and the derivatives were approximated by the difference ratios. 

The positions of maxima are listed in Table\ \ref{mtrgb}. For comparison, we also
show the dips of the LFs and the TRGB values determined by Cioni et al. (2000). 
The overall agreement is excellent: most of the differences do not exceed the
uncertainties. The largest discrepancy was found in $K_{\rm S}$, which can,
however, be explained by a $\sim$0.1 mag systematic shift between 2MASS  and
DENIS $K$-magnitudes. Therefore, we conclude the luminosity functions of OGLE-II
variables show exactly the same behaviour as, for instance, those of the much 
larger DENIS Catalogue towards the Magellanic Clouds, which included 
both variable and non-variable stars.

Additionally, we find a significant wavelength
dependence of the TRGB values, which follows 
exactly the changes observed in the tip of the RGB as a function
of metallicity in globular clusters by Ferraro et al. (2000) and Ivanov \&
Borissova (2002). For instance, Ferraro et al. (2000) derived 
$M_{\rm K}^{\rm tip}\sim-0.6$[Fe/H], which implies $\Delta K\approx0.18$ mag,
assuming the metallicity differs by a factor of 2 between the Clouds. The
observed differences from the second derivatives 
are 0.19 mag, 0.25 mag and 0.23 mag in $J$, $H$ and
$K_{\rm S}$, respectively (adopting $\Delta \mu_0=0.44$ mag and correcting for
differential extinctions) and they are 
also in good agreement with the relations in Fig.\ 4 of Ivanov \& Borissova
(2002). In $I$ band, the difference is only 0.03 mag, which illustrates
the insensitivity of $M_{\rm TRGB}(I)$ to metallicity and age (Lee et al. 1993).

We interpret these agreements as our last piece of evidence that the dip between 
the AGB and RGB peaks indeed corresponds to the tip of the RGB. In Paper I, we
have also considered the arguments of Alves et al. (1998) and Wood (2000) 
that all variables below the TRGB are thermally pulsing AGB stars and that their
brighter cut-off matches the TRGB by coincidence. With the presented behaviour,
this explanation seems to be quite unlikely. Even if there are faint TP-AGB
stars below the tip, their fraction must be small compared to first ascent red
giants. 

\subsection{Amplitude distributions}

We finish the comparative analysis by presenting  the same six slices 
of the (period, amplitude, $K_{\rm S}$ magnitude) data cube as were shown for the
LMC in Fig.\ 4 of Paper I. Colours refer  to three
$J-K_{\rm S}$ colour ranges, enabling a rough classification of stars as ``hot'',
``warm'' and  ``cool'' red giants.

The distributions are similar to those of the LMC in that there is a good
correlation between the amplitude and mode of pulsation. For amplitudes larger
than 0.04 mag, practically all variables below the TRGB disappear. Furthermore,
the colour distribution within any particular P-L ridge follows the same pattern
as in the LMC. There are, however, some striking differences. Firstly, there is a
relative lack of ``warm'' stars among the low-amplitude variables of the SMC.
There is one blue dot for every  30 dots in the upper left panel; in the LMC, the
ratio is 1:10. A similar  under-abundance is also visible in the lower panels.
Secondly, the lower three panels are dominated by the very red stars with
$J-K_{\rm S}>1.4$ mag, which are probably carbon stars.  For the highest
amplitude variables (all with regular Mira-like light curves), there is
only a handful  of ``hot'' and ``warm'' giants, many of which fall below  the
fiducial Mira P-L by as much as 1.5--2 mag. These ``faint Miras'' be 
identified with the
dust-enshrouded Mira stars (Wood 1998), since their extinctions increase strongly
toward shorter wavelengths, reaching up to 4-6 magnitudes in $I$. Our data
illustrate nicely that, despite the fact that carbon-rich Mira stars obey the
same P-L relation as the oxygen-rich Miras (Feast et al. 1989), the use  of the
Mira P-L relation in extragalactic distance measurements must be restricted to
the bluest possible Mira stars.

\section{Conclusions}
 
In this paper we presented the results of a period analysis of OGLE-II red variable
stars in the SMC. Multiple periodicity has been found for the majority of stars,
and can now be identified as common behaviour in pulsating red giants,
regardless of the metallicity. 

We confirm the existence of distinct P-L relations below
the tip of the Red Giant Branch, announced in Paper I for the LMC. Utilizing  the
overall P-L distributions, we determined multi-wavelength relative distance  moduli
and extinctions, which support a consistent view that the differential 
distance modulus of the Clouds is $\Delta
\mu_0\approx$\ 0.44 mag. The luminosity functions revealed the AGB variables
follow very similar distributions in both Clouds. We showed
that the dips between the AGB and supposed RGB peaks do indeed correspond to the
tip of the RGB (within a few hundredths of a mag). This is also supported by the
wavelength-dependent $\Delta M_{\rm TRGB}$ ($\sim$0.20 mag in $JHK_{\rm S}$)
found as an excess to the relative distance modulus. It is
consistent with the metallicity dependence seen among globular clusters and we
take this as further evidence that the contamination by thermally pulsing AGB
stars is likely to be small. 

\section*{Acknowledgments} 

This work has been supported by the FKFP Grant
0010/2001, OTKA Grant \#F043203 and the Australian Research Council. Thanks are 
due to an anonymous referee, whose suggestions led to significant improvement of
the paper. This
research has made use of the NASA/IPAC Infrared Science Archive, which is
operated by the Jet Propulsion Laboratory, California Institute of Technology,
under contract with the National Aeronautics and Space Administration. 
The NASA ADS Abstract Service was used to access data and references.

\end{document}